\documentclass[12pt]{article}
\usepackage{mathrsfs}
\usepackage{amssymb}
\usepackage{graphics}
\usepackage{graphicx}
\usepackage{ulem}%under wave package
\renewcommand{\theequation}{\arabic{section}.\arabic{equation}}

\begin{document}

%************************** Text Begins here ******************************

%  Greek letters

\def\a{\alpha}
\def\b{\beta}
\def\d{\delta}
\def\e{\epsilon}
\def\g{\gamma}
    \def\h{\mathfrak{h}}
\def\k{\kappa}
\def\l{\lambda}
\def\o{\omega}
\def\p{\wp}
\def\r{\rho}
\def\t{\tau}
\def\s{\sigma}
\def\z{\zeta}
\def\x{\xi}
 \def\A{{\cal{A}}}
 \def\B{{\cal{B}}}
 \def\C{{\cal{C}}}
 \def\D{{\cal{D}}}
\def\G{\Gamma}
\def\K{{\cal{K}}}
\def\O{\Omega}
\def\R{\bar{R}}
\def\T{{\cal{T}}}
\def\L{\Lambda}
\def\f{E_{\tau,\eta}(sl_2)}
\def\E{E_{\tau,\eta}(sl_n)}
\def\Zb{\mathbb{Z}}
\def\Cb{\mathbb{C}}

\def\R{\overline{R}}
% Shorthands for \begin{equation} and the like

\def\beq{\begin{equation}}
\def\eeq{\end{equation}}
\def\bea{\begin{eqnarray}}
\def\eea{\end{eqnarray}}
\def\ba{\begin{array}}
\def\ea{\end{array}}
\def\no{\nonumber}
\def\le{\langle}
\def\re{\rangle}
\def\lt{\left}
\def\rt{\right}

\newtheorem{Theorem}{Theorem}
\newtheorem{Definition}{Definition}
\newtheorem{Proposition}{Proposition}
\newtheorem{Lemma}{Lemma}
\newtheorem{Corollary}{Corollary}
\newcommand{\proof}[1]{{\bf Proof. }
        #1\begin{flushright}$\Box$\end{flushright}}

\baselineskip=20pt

%%%%%%%%%%%%%%%%%%%%%%%%%%%%%%%%%%%%%%%%%%%%%%%%%%%%%%%%%%%%
%                                                          %
%  Title page                                              %
%                                                          %
%%%%%%%%%%%%%%%%%%%%%%%%%%%%%%%%%%%%%%%%%%%%%%%%%%%%%%%%%%%%
\newfont{\elevenmib}{cmmib10 scaled\magstep1}
\newcommand{\preprint}{
   \begin{flushleft}
     %\elevenmib Yukawa\, Institute\, Kyoto\\
   \end{flushleft}\vspace{-1.3cm}
   \begin{flushright}\normalsize
  % \sf  YITP-03-53\\
  %  {\tt hep-th/0703222} \\ March 2007
   \end{flushright}}
\newcommand{\Title}[1]{{\baselineskip=26pt
   \begin{center} \Large \bf #1 \\ \ \\ \end{center}}}
\newcommand{\Author}{\begin{center}
   \large \bf
Kun Hao\,${}^{a}$, {Wen-Li Yang${}^{a}$\,\footnote{Corresponding author: wlyang@nwu.edu.cn (W.-L. Yang).}}, Heng Fan\,${}^b$, Si-Yuan Liu\,${}^a$,
Ke Wu\,${}^c$, Zhan-Ying Yang\,${}^d$  and Yao-Zhong Zhang\,${}^e$

 \end{center}}
\newcommand{\Address}{\begin{center}

     ${}^a$ Institute of Modern Physics, Northwest University,
     Xian 710069, P.R. China\\
     ${}^b$ Beijing National Laboratory for Condensed Matter Physics,
     Institute of Physics, \\ Chinese Academy of Sciences, Beijing 100190, P.R. China\\
     ${}^c$ School of Mathematical Science, Capital Normal University,
     Beijing 100037, P.R. China \\
     ${}^d$ The Department of Physics, Northwest University,
     Xian 710069, P.R. China \\
     ${}^e$ School of Mathematics and Physics, The University of Queensland, Brisbane,
       QLD 4072, Australia\\
E-mail: haoke72@163.com, wlyang@nwu.edu.cn, hfan@aphy.iphy.ac.cn, lsy5227@163.com, wuke@mail.cnu.edu.cn, zyyang@nwu.edu.cn, yzz@maths.uq.edu.au

   \end{center}}
\newcommand{\Accepted}[1]{\begin{center}
   {\large \sf #1}\\ \vspace{1mm}{\small \sf Accepted for Publication}
   \end{center}}

\preprint
\thispagestyle{empty}
\bigskip\bigskip\bigskip

\Title{
Determinant representations for scalar products of the XXZ Gaudin model with general boundary terms} \Author

\Address
\vspace{1cm}

\begin{abstract}
We obtain the determinant representations of the scalar products for the XXZ Gaudin model with generic non-diagonal boundary terms.

\vspace{1truecm} \noindent {\it PACS:} 03.65.Fd; 04.20.Jb;
05.30.-d; 75.10.Jm

\end{abstract}
\newpage
%%%%%%%%%%%%%%%%%%%%%%%%%%%%%%%%%%%%%%%%%%%%%%%%%%%%%%%%%%%%%%%
%                                                             %
%  1. Introduction                                            %
%                                                             %
%%%%%%%%%%%%%%%%%%%%%%%%%%%%%%%%%%%%%%%%%%%%%%%%%%%%%%%%%%%%%%%
\section{Introduction}
\label{intro} \setcounter{equation}{0}
Gaudin type models \cite{Gau76} have important applications in physics. For instance, the XXZ Gaudin model has played an essential role
in the study of the reduced BCS model whose exact solutions were first found by Richardson \cite{Richardson1963}.
In fact, the conserved operators for the BCS model can be mapped to a set of XXX Gaudin Hamiltonians
in a non-uniform magnetic field and the BCS Hamiltonian is expressible in terms of these operators \cite{Lor02,Sierra00,Duk04}.
This result also exposes a relationship between the BCS Hamiltonian and the perturbed WZNW model at critical level \cite{Fei94}
based on the connection of Gaudin models  with the solutions of the KZ equations \cite{Babujian93,Reshetikhin95}.

Eigenstates and the corresponding eigenvalues for the open XXZ Gaudin model for the boundary conditions with three
free boundary parameters were derived in Ref.\cite{Yang04a}. In this paper, we consider the most generic boundary conditions
specified by the non-diagonal K-matrices in \cite{Veg93,Gho94}, leading to the open XXZ Gaudin model in this paper which depends on
four free boundary parameters. We compute the scalar products of this Gaudin model, and give their explicit expressions in terms of determinants,
that is the determinant representations of the scalar products.

This paper is organized as follows. Section 2 provides some preliminaries on the boundary inverse scattering method.
In section 3, we briefly describe the open XXZ Gaudin magnet associated with non-diagonal boundary K-matrices. In section 4, we
derive the Bethe ansatz equations for the open XXZ Gaudin model and
the symmetric, polarization-free expressions for the pseudo-particle creation operators.
In section 5, we obtain the determinant representations for the partition and correlation functions of the model.
We summarize our results in section 6 and present the details of some derivations and proofs in the Appendices.

%%%%%%%%%%%%%%%%%%%%%%%%%%%%%%%%%%%%%%%%%%%%%%%%%%%%%%%%%%%%%%%
%                                                             %
%  2. The inhomogeneous spin-$\frac{1}{2}$                    %
%                    XXZ open chain                           %
%                                                             %
%                                                             %
%%%%%%%%%%%%%%%%%%%%%%%%%%%%%%%%%%%%%%%%%%%%%%%%%%%%%%%%%%%%%%%

\section{ Preliminaries: the inhomogeneous spin-$\frac{1}{2}$ XXZ open chain}
\label{XXZ} \setcounter{equation}{0}

Let $V$ be a two-dimensional linear space and $\s^{\pm}$, $\s^z$ be the Pauli matrices which give the spin-$\frac{1}{2}$
representation of $su(2)$ on V. The
spin-$\frac{1}{2}$ XXZ chain can be constructed from the
well-known six-vertex model R-matrix $R(u)\in {\rm End}(V\otimes
V)$ \cite{Kor93} given by \bea
R(u)=\lt(\begin{array}{llll}1&&&\\&\frac{\sin u}{\sin(u+\eta)}&\frac{\sin\eta}{\sin(u+\eta)}&\\
&\frac{\sin\eta}{\sin(u+\eta)}&\frac{\sin u}{\sin(u+\eta)}&\\&&&1\end{array}\rt).\label{r-matrix}\eea
Here we assume  $\eta$ is a
generic complex number. The R-matrix satisfies the quantum
Yang-Baxter equation (QYBE), \bea
R_{1,2}(u_1-u_2)R_{1,3}(u_1-u_3)R_{2,3}(u_2-u_3)
=R_{2,3}(u_2-u_3)R_{1,3}(u_1-u_3)R_{1,2}(u_1-u_2),\label{QYB}\eea
and the unitarity, crossing-unitarity and quasi-classical
properties \cite{Yang04a}. We adopt the standard notations: for
any matrix $A\in {\rm End}(V)$ , $A_j$ is an embedding
operator in the tensor space $V\otimes V\otimes\cdots$, which acts
as $A$ on the $j$-th space and as identity on the other factor
spaces; $R_{i,j}(u)$ is an embedding operator of R-matrix in the
tensor space, which acts as identity on the factor spaces except
for the $i$-th and $j$-th ones.

One introduces the ``row-to-row"  (or one-row) monodromy matrix
$T(u)$, which is an $2\times 2$ matrix with elements being
operators acting on $V^{\otimes N}$, where $N=2M$ ($M$ being a
positive integer),\bea
T_0(u)=R_{0,N}(u-z_N)R_{0,N-1}(u-z_{N-1})\cdots
R_{0,1}(u-z_1).\label{Mon-V}\eea Here $\{z_j|j=1,\cdots,N\}$ are
arbitrary free complex parameters which are usually called
inhomogeneous parameters.

Integrable open chain can be constructed as follows \cite{Skl88}.
Let us introduce a pair of K-matrices $K^-(u)$ and $K^+(u)$. The
former satisfies the reflection equation (RE)
 \bea &&R_{1,2}(u_1-u_2)K^-_1(u_1)R_{2,1}(u_1+u_2)K^-_2(u_2)\no\\
 &&~~~~~~=
K^-_2(u_2)R_{1,2}(u_1+u_2)K^-_1(u_1)R_{2,1}(u_1-u_2),\label{RE-V}\eea
and the latter  satisfies the dual RE \bea
&&R_{1,2}(u_2-u_1)K^+_1(u_1)R_{2,1}(-u_1-u_2-2\eta)K^+_2(u_2)\no\\
&&~~~~~~=
K^+_2(u_2)R_{1,2}(-u_1-u_2-2\eta)K^+_1(u_1)R_{2,1}(u_2-u_1).
\label{DRE-V}\eea For open spin-chains, instead of the standard
``row-to-row" monodromy matrix $T(u)$ (\ref{Mon-V}), one needs to
consider  the
 ``double-row" monodromy matrix $\mathbb{T}(u)$
\bea
  \mathbb{T}(u)=T(u)K^-(u)\hat{T}(u),\quad \hat{T}(u)=T^{-1}(-u).
  \label{Mon-V-0}
\eea Then the double-row transfer matrix of the XXZ chain with
open boundary (or the open XXZ chain) is given by \bea
\t(u)=tr(K^+(u)\mathbb{T}(u)).\label{trans}\eea The QYBE and
(dual) REs lead to that the transfer matrices with different
spectral parameters commute with each other \cite{Skl88}:
$[\t(u),\t(v)]=0$. This ensures the integrability of the open XXZ
chain.

%%%%%%%%%%%%%%%%%%%%%%%%%%%%%%%%%%%%%%%%%%%%%%%%%%%%%%%%%%%%%%%
%                                                             %
%  3. XXZ Gaudin model with generic boundaries                %
%                                                             %
%                                                             %
%%%%%%%%%%%%%%%%%%%%%%%%%%%%%%%%%%%%%%%%%%%%%%%%%%%%%%%%%%%%%%%

\section{ XXZ Gaudin model with generic boundaries}
\label{Gaudin} \setcounter{equation}{0}

We will consider the K-matrix $K^{-}(u)$ which is a
generic solution to the RE (\ref{RE-V}) associated the six-vertex
model R-matrix  \cite{Veg93,Gho94}
\bea K^-(u)=\lt(\begin{array}{ll}k_1^1(u)&k^1_2(u)\\
k^2_1(u)&k^2_2(u)\end{array}\rt)\equiv K(u).\label{K-matrix}
\eea
The matrix elements are \bea && k^1_1(u)=
\frac{\cos(\l_1-\l_2) -\cos(\l_1+\l_2+2\xi)e^{-2iu}}
{2\sin(\l_1+\xi+u)
\sin(\l_2+\xi+u)},\no\\
&&k^1_2(u)=\frac{-i\sin(2u)e^{-i(\l_1+\l_2)} e^{-iu}}
{2\sin(\l_1+\xi+u) \sin(\l_2+\xi+u)},\no\\
&&k^2_1(u)=\frac{i\sin(2u)e^{i(\l_1+\l_2)} e^{-iu}}
{2\sin(\l_1+\xi+u) \sin(\l_2+\xi+u)}, \no\\
&& k^2_2(u)=\frac{\cos(\l_1-\l_2)e^{-2iu}- \cos(\l_1+\l_2+2\xi)}
{2\sin(\l_1+\xi+u)\sin(\l_2+\xi+u)}.\label{K-matrix-2-1}
\eea
The corresponding {\it dual\/} K-matrix $K^+(u)$ is a generic solution to the dual
reflection equation (\ref{DRE-V}) with a particular choice of the
free boundary parameters:
\bea K^+(u)=\lt(\begin{array}{ll}{k^+}_1^1(u)&{k^+}^1_2(u)\\
{k^+}^2_1(u)&{k^+}^2_2(u)\end{array}\rt)\label{DK-matrix}
\eea
with matrix elements
\bea && {k^+}^1_1(u)=
\frac{\cos(\l_1-\l_2)e^{-i\eta}
-\cos(\l_1+\l_2+2\bar{\xi})e^{2iu+i\eta}}
{2\sin(\l_1+\bar{\xi}-u-\eta)
\sin(\l_2+\bar{\xi}-u-\eta)},\no\\
&&{k^+}^1_2(u)=\frac{i\sin(2u+2\eta)e^{-i(\l_1+\l_2)}
e^{iu-i\eta}} {2\sin(\l_1+\bar{\xi}-u-\eta)
\sin(\l_2+\bar{\xi}-u-\eta)},
\no\\
&&{k^+}^2_1(u)=\frac{-i\sin(2u+2\eta)e^{i(\l_1+\l_2)}
e^{iu+i\eta}}
{2\sin(\l_1+\bar{\xi}-u-\eta) \sin(\l_2+\bar{\xi}-u-\eta)}, \no\\
&& {k^+}^2_2(u)=\frac{\cos(\l_1-\l_2)e^{2iu+i\eta}-
\cos(\l_1+\l_2+2\bar{\xi})e^{-i\eta}}
{2\sin(\l_1+\bar{\xi}-u-\eta)\sin(\l_2+\bar{\xi}-u-\eta)}.\label{K-matrix-6}
\eea The K-matrices depend on four free boundary parameters
$\{\l_1,\,\l_2,\,\xi,\,\bar{\xi}\}$ which specify integrable boundary conditions \cite{Gho94}.
We remark that $K^-(u)$ does not depend
on the crossing parameter $\eta$ but $K^+(u)$ does. The parameter $\bar{\xi}$ is required to have the following expansion:
\bea
\bar{\xi}=\xi+\eta \Delta+O(\eta^2),\quad\quad \eta\rightarrow 0.
\label{Delta}\eea
This results in the relation,
 \bea \lim_{\eta\rightarrow
0}\{K^+(u)K^-(u)\}=\lim_{\eta\rightarrow 0}\{K^+(u)\}K(u)={\rm id}.\label{ID-1}
\eea
Let us introduce the generalized  XXZ Gaudin operators
\cite{Gau76} $\{H_j |j=1,2,\cdots,N\}$ associated with the
spin-$\frac{1}{2}$ XXZ  model with boundaries specified by
the K-matrices (\ref{K-matrix}) and (\ref{DK-matrix}):
\bea &&H_j=\G_j(z_j)+\sum_{k\neq
j}^{2M}\frac{1}{\sin(z_j-z_k)}\lt\{\s^+_k\s^-_j+\s^-_k\s^+_j
+\cos(z_j-z_k)\frac{\s^z_k\s^z_j-1}{2}\rt\}\no\\
&&~~+\sum_{k\neq
j}^{2M}\frac{K_j^{-1}(z_j)}{\sin(z_j+z_k)}\lt\{\s^+_j\s^-_k+\s^-_j\s^+_k
+\cos(z_j+z_k)\frac{\s^z_j\s^z_k-1}{2}\rt\}K_j(z_j),\label{Ham}
\eea  where $\G_j(u)=\frac{\partial}{\partial
\eta}\{\bar{K}_j(u)\}|_{\eta=0}K_j(u)$, $j=1,\cdots,N,$ with
$\bar{K}_j(u)=tr_0\lt\{K^+_0(u)R_{0j}(2u)P_{0j}\rt\}$, and
$\{z_j\}$ correspond to the inhomogeneous parameters of the
spin-$\frac{1}{2}$ XXZ chain with generic open
boundaries. For a generic choice of the boundary parameters
$\{\l_1,\,\l_2,\,\xi,\bar{\xi}\}$, $\G_j(u)$ is an non-diagonal matrix, in
contrast to the situation in \cite{Lor02}.

The XXZ Gaudin operators (\ref{Ham}) can be obtained by expanding the
double-row transfer matrix $\t(u)$ (\ref{trans}) at
$u=z_j$ around $\eta=0$ \cite{Yang04a}: \bea &&\t(z_j)={\rm id}+\eta
H_j+O(\eta^2),~~j=1,\cdots,N, \label{trans-2}\\
&&H_j=\frac{\partial}{\partial
\eta}\t(z_j)|_{\eta=0}.\label{Eq-1}
\eea Then the commutativity of the transfer matrices $\{\t(z_j)\}$
for a generic $\eta$ implies that
\bea
[H_j,H_k]=0,~~i,j=1,\cdots,N.\label{Com-1}
\eea Thus the Gaudin model with the local Hamiltonian (\ref{Ham}) is integrable.

%%%%%%%%%%%%%%%%%%%%%%%%%%%%%%%%%%%%%%%%%%%%%%%%%%%%%%%%%%%%%%%%%%%%%%%%%%%%%%%%

%%%%%%%%%%%%%%%%%%%%%%%%%%%%%%%%%%%%%%%%%%%%%%%%%%%%%%%%%%%%%%%
%                                                             %
%  4. Eigenstates and the corresponding eigenvalues           %
%                                                             %
%                                                             %
%%%%%%%%%%%%%%%%%%%%%%%%%%%%%%%%%%%%%%%%%%%%%%%%%%%%%%%%%%%%%%%

\section{ Eigenstates and the corresponding eigenvalues}
\label{BAsection} \setcounter{equation}{0}

The relation (\ref{Eq-1}) between $\{H_j\}$ and $\{\t(z_j)\}$ and
the fact that the first term of (\ref{trans-2}) is the identity operator
enable us to extract the eigenstates of the Gaudin operators and
the corresponding eigenvalues from those of the XXZ chain  obtained in \cite{Nep04,Cao03,Yang04a,Gal05,Gie05,Gie05-1,Baj06,Doi06,Mur06,Yang07,Yang11}.

Let us introduce the states $|\O^{(1)}\rangle$ and $|\O^{(2)}\rangle$,
\bea
  &&|\O^{(1)}\rangle=\lt(\begin{array}{c}e^{-i(z_1+2\l_1)}\\1\end{array}\rt)
      \otimes\cdots\otimes
      \lt(\begin{array}{c}e^{-i(z_N+2\l_1)}\\1\end{array}\rt),\label{Vac-1}\\[6pt]
  &&|\O^{(2)}\rangle=\lt(\begin{array}{c}e^{-i(z_1+2\l_2)}\\1\end{array}\rt)
      \otimes\cdots\otimes
      \lt(\begin{array}{c}e^{-i(z_N+2\l_2)}\\1\end{array}\rt),\label{Vac-2}
\eea and the matrix $g(u)\in {\rm End}(V)$ and the associated gauged Pauli operator $\s^{\pm}(u)\in {\rm
End}(V)$
\bea
  g(u)&=&\lt(\begin{array}{cc}e^{-i(u+2\l_1)}&e^{-i(u+2\l_2)}
       \\1&1\end{array}\rt),\label{Matrix-in}\\[6pt]
  \s^{\pm}(u)&=&g(u)\s^{\pm}g(u)^{-1}.\label{Matrix-in-1}
\eea Then we define the states,
\bea
&&|\{v_i^{(1)}\}\rangle^{(1)}=\prod_{i=1}^{M}
B(v_i^{(1)})|\O^{(1)}\rangle,
   \label{Bethe-state-1}\\
&&|\{v_i^{(2)}\}\rangle^{(2)} =\prod_{i=1}^{M}
C(v_i^{(2)})|\O^{(2)}\rangle.
   \label{Bethe-state-2}
\eea The associated operators $B(u)$ and $C(u)$ are
\bea
 B(u)&=&
 \sum^{N}_{i=1}\frac{\sin(\l_1+\x+z_i)\sin(\l_2+\x-z_i)\cdot\sin(2u)}
 {\sin(\l_1+\x-u)\sin(\l_2+\x-u)\sin(u-z_i)\sin(u+z_i)}
 \times \s^-(z_i),\label{Operator-B}\\
 C(u)&=&\sum^{N}_{i=1}\frac{\sin(\l_1+\x-z_i)\sin(\l_2+\x+z_i)\cdot\sin(2u)}
 {\sin(\l_1+\x+u)\sin(\l_2+\x+u)\sin(u-z_i)\sin(u+z_i)}
 \times \s^+(z_i).\label{Operator-C}
 \eea Using the same method as in \cite{Hik95}, we can show that the above states (\ref{Bethe-state-1}) and
 (\ref{Bethe-state-2}) are the common eigenstates of the Gaudin operators $\{H_j\}$ given by (\ref{Ham}) provided that
 the parameters $\{v_i^{(i)}\}$ satisfy the following two sets of Bethe ansatz equations
\bea
&&\frac{1-\Delta}{\sin(\l_1+\x+v^{(1)}_{\a})\sin(\l_1+\x-v^{(1)}_{\a})}
+\frac{1+\Delta}{\sin(\l_2+\x+v^{(1)}_{\a})\sin(\l_2+\x-v^{(1)}_{\a})}\no\\
&=&\sum_{k\neq\a}^M\frac{2}{\sin(v^{(1)}_{\a}-v^{(1)}_k)\sin(v^{(1)}_{\a}+v^{(1)}_k)}
-\sum_{k=1}^{2M}\frac{1}{\sin(v^{(1)}_{\a}-z_k)\sin(v^{(1)}_{\a}+z_k)},\label{BA}\\
%&&~~\a=1,\cdots,M,
%\label{BA}
%\eea
%or
%\bea
&&\frac{1+\Delta}{\sin(\l_1+\x+v^{(2)}_{\a})\sin(\l_1+\x-v^{(2)}_{\a})}
+\frac{1-\Delta}{\sin(\l_2+\x+v^{(2)}_{\a})\sin(\l_2+\x-v^{(2)}_{\a})}\no\\
&=&\sum_{k\neq\a}^M\frac{2}{\sin(v^{(2)}_{\a}-v^{(2)}_k)\sin(v^{(2)}_{\a}+v^{(2)}_k)}
-\sum_{k=1}^{2M}\frac{1}{\sin(v^{(2)}_{\a}-z_k)\sin(v^{(2)}_{\a}+z_k)},\no\\
&&~~~~~~~~~~~~\a=1,\cdots,M. \label{BA2}
\eea
Here $\Delta$ is the parameter of first order expansion of $\bar{\xi}$ in terms of $\eta$, as defined in (\ref{Delta}).
Namely,
\bea
  H_j\,|\{v_\a^{(i)}\}\rangle^{(i)}=E_j^{(i)}|\{v_\a^{(i)}\}\rangle^{(i)},\qquad i=1,2,
\eea
where $E_j^{(i)}$ are given by
\bea
E_j^{(1)}&=&\cot 2z_j+\sum_{j=1}^2\cot(\l_j+\xi-z_j)
   -\frac{\Delta\sin(2z_j)}{\sin(\l_1+\xi-z_j)\sin(\l_1+\xi+z_j)}\no\\
   &&+\sum_{k=1}^M
       \frac{\sin (2z_j)}{\sin(v_k^{(1)}-z_j)\sin(v_k^{(1)}+z_j)},\label{Eig-1}
       \\
E_j^{(2)}&=&\cot 2z_j+\sum_{j=1}^2\cot(\l_j+\xi-z_j)
   -\frac{\Delta\sin(2z_j)}{\sin(\l_2+\xi-z_j)\sin(\l_2+\xi+z_j)}\no\\
&&   +\sum_{k=1}^M
       \frac{\sin (2z_j)}{\sin(v_k^{(2)}-z_j)\sin(v_k^{(2)}+z_j)}.\label{Eig-2}
\eea

\section{Determinant representations of the scalar products}
\label{Scalar Products} \setcounter{equation}{0}

To obtain correlation functions, it suffices to calculate the scalar products
of on-shell Bethe states with general off-shell Bethe states  \cite{Kor93} (see also
\cite{Wan02,Kit07} for the open XXZ chain with diagonal boundaries).
In this section, we will obtain the explicit expressions of the following scalar products
for the open XXZ Gaudin model with non-diagonal boundary terms:
\bea
&&\hspace{-1.2truecm}S^{1,2}(\{u_{\a}\};\{v_i^{(2)}\})=
   {}^{(1)}\langle\{u_{\a}\}|\{v_i^{(2)}\}\rangle^{(2)},\quad
   S^{2,1}(\{u_{\a}\};\{v_i^{(1)}\})= {}^{(2)}\langle\{u_{\a}\}
   |\{v_i^{(1)}\}\rangle^{(1)},\label{Scalar-1}\\
&&\hspace{-1.2truecm}S^{1,1}(\{u_{\a}\};\{v_i^{(1)}\})=
   {}^{(1)}\langle\{u_{\a}\}|\{v_i^{(1)}\}\rangle^{(1)},\quad
   S^{2,2}(\{u_{\a}\};\{v_i^{(2)}\})=
   {}^{(2)}\langle\{u_{\a}\}|\{v_i^{(2)}\}\rangle^{(2)},\label{Scalar-2}
\eea where ${}^{(1)}\langle\{u_{\a}\}|$ and
${}^{(2)}\langle\{u_{\a}\}|$ are defined by
\bea
&&{}^{(1)}\langle\{u_{\a}\}|=\langle\Omega^{(1)}|C(u_M)\ldots C(u_1),\label{Dual-1}\\
&&{}^{(2)}\langle\{u_{\a}\}|=\langle\Omega^{(2)}|B(u_M)\ldots B(u_1),\label{Dual-2}
\eea
with $\langle\Omega^{(1)}|$, $\langle\Omega^{(2)}|$ being the dual states of $|\O^{(1)}\rangle$, $|\O^{(2)}\rangle$,
respectively,
\bea
   &&\langle\Omega^{(1)}|\hspace{-0.12truecm}=\hspace{-0.12truecm}\lt\{\prod_{j=1}^{N}\frac{ie^{-i(z_j+\l_1+\l_2)}}{2\sin(\l_1-\l_2)}\rt\}
       \lt(\begin{array}{cc}1,&-e^{-i(z_1+2\l_2)}\end{array}\rt)
      \otimes\cdots\otimes
      \lt(\begin{array}{cc}1,&-e^{-i(z_N+2\l_2)}\end{array}\rt),\label{Vac-3}\\
   &&\langle\Omega^{(2)}|\hspace{-0.12truecm}=\hspace{-0.12truecm}\lt\{\prod_{j=1}^{N}\frac{ie^{-i(z_j+\l_1+\l_2)}}{2\sin(\l_1-\l_2)}\rt\}
       \lt(\begin{array}{cc}-1,&e^{-i(z_1+2\l_1)}\end{array}\rt)
      \otimes\cdots\otimes
      \lt(\begin{array}{cc}-1,&e^{-i(z_N+2\l_1)}\end{array}\rt).\label{Vac-4}
\eea

By means of (\ref{Vac-1})-(\ref{Operator-C}), (\ref{Vac-3}) and (\ref{Vac-4}), the scalar products can be
written as
\bea
   S^{1,2}(\{u_{\a}\};\{v_i\})&=&\langle \Uparrow|\tilde{C}(u_1)\cdots \tilde{C}(u_M)\,\tilde{C}(v_1)\cdots
        \tilde{C}(v_M)|\Downarrow\rangle,\\
   S^{2,1}(\{u_{\a}\};\{v_i\})&=&\langle \Downarrow|\tilde{B}(u_1)\cdots \tilde{B}(u_M)\,\tilde{B}(v_1)\cdots
        \tilde{B}(v_M)|\Uparrow\rangle,\\
   S^{1,1}(\{u_{\a}\};\{v_i\})&=& \langle \Uparrow|\tilde{C}(u_1)\cdots \tilde{C}(u_M)  \,\tilde{B}(v_1)\cdots
        \tilde{B}(v_M)|\Uparrow\rangle,\label{ScalarBethe1}\\
   S^{2,2}(\{u_{\a}\};\{v_i\})&=& \langle \Downarrow|\tilde{B}(u_1)\cdots \tilde{B}(u_M)\, \tilde{C}(v_1)\cdots
        \tilde{C}(v_M)|\Downarrow\rangle.\label{ScalarBethe2}
\eea
Here $|\Uparrow\rangle$ and $\langle \Uparrow|$ (resp. $|\Downarrow\rangle$ and $\langle \Downarrow|$) are the all spin-up state
and its dual (resp. all spin-down and its dual), and $\tilde{C}(u)$ and $\tilde{B}(u)$ are given by
\bea
 \tilde{B}(u)&=&
 \sum^{N}_{i=1}\frac{\sin(\l_1+\x+z_i)\sin(\l_2+\x-z_i)\sin(2u)}
 {\sin(\l_1+\x-u)\sin(\l_2+\x-u)\sin(u-z_i)\sin(u+z_i)}
 \times \s^-_i,\label{T+12}\\
 \tilde{C}(u)&=&\sum^{N}_{i=1}\frac{\sin(\l_1+\x-z_i)\sin(\l_2+\x+z_i)\sin(2u)}
 {\sin(\l_1+\x+u)\sin(\l_2+\x+u)\sin(u-z_i)\sin(u+z_i)}
 \times \s^+_i.\label{T-21}
 \eea

%%%%%%%%%%%%%%%%%%%%%%%%%%%%%%%%%%%%%%%%%%%%%%%%%%%%%%%%%%%%%%%%%%%%%%%%%%%
\subsection{Scalar products $S^{1,2}$ and $S^{2,1}$}
Let us introduce two functions
\bea
 Z^{(1)}_N(\{\bar{u}_{J}\})&\equiv& S^{1,2}(\{u_{\a}\};\{v_i\})=
     \langle \Uparrow|\tilde{C}(u_1)\cdots \tilde{C}(u_M)\,\tilde{C}(v_1)\cdots
     \tilde{C}(v_M)|\Downarrow\rangle,\\
 Z^{(2)}_N(\{\bar{u}_{J}\})&\equiv& S^{2,1}(\{u_{\a}\};\{v_i\})=
     \langle \Downarrow|\tilde{B}(u_1)\cdots \tilde{B}(u_M)\,\tilde{B}(v_1)\cdots
     \tilde{B}(v_M)|\Uparrow\rangle,
\eea where the $N$ parameters $\{\bar{u}_J|J=1,\ldots N\}$ are defined as
\bea
\bar{u}_i=u_i\,\, {\rm for}\,\, i=1,\ldots M,\qquad {\rm and}\qquad\bar{u}_{M+i}=v_i \,\, {\rm for}\,\, i=1,\ldots M.
\eea
Note that these functions $Z^{(1)}_N(\{\bar{u}_{J}\})$ and $Z^{(2)}_N(\{\bar{u}_{J}\})$ may correspond to the partition functions
of the Gaudin model with domain wall boundary condition and one reflecting end \cite{Tsu98,Yang10} specified by the non-diagonal
 K-matrices (\ref{K-matrix}) and (\ref{DK-matrix}).

We find that the functions $Z^{(k)}_N(\{\bar{u}_{J}\})$ above can be expressed in term of the determinants
of the $N\times{N}$ matrices $\mathscr{N}^{(k)}(\{\bar{u}_\a\};\{z_i\})_{\a,j}$,
\bea
Z^{(k)}_N(\{\bar{u}_j\}) =\frac{\prod^{N}_{\a=1}\prod^{N}_{i=1}\sin(\bar{u}_{\a}+z_i)\sin(\bar{u}_{\a}-z_i)
det\mathscr{N}^{(k)}(\{\bar{u}_\a\};\{z_i\})_{\a,j}}
{\prod_{\a>\b}\sin(\bar{u}_{\a}-\bar{u}_{\b})
\sin(\bar{u}_{\a}+\bar{u}_{\b})\prod_{k<j}\sin(z_k-z_j)\sin(z_k+z_j)}, \label{partition function}
\eea
where
\bea
\mathscr{N}^{(1)}(\{\bar{u}_\a\};\{z_i\})_{\a,j}=
\frac{\sin(\l_1+\x-z_i)\sin(\l_2+\x+z_i)\sin(2\bar{u})}
{\sin(\l_1+\x+\bar{u})\sin(\l_2+\x+\bar{u})\sin^2(\bar{u}-z_i)\sin^2(\bar{u}+z_i)},
\eea
%and
\bea
\mathscr{N}^{(2)}(\{\bar{u}_\a\};\{z_i\})_{\a,j}=
\frac{\sin(\l_1+\x+z_i)\sin(\l_2+\x-z_i)\sin(2\bar{u})}
{\sin(\l_1+\x-\bar{u})\sin(\l_2+\x-\bar{u})\sin^2(\bar{u}-z_i)\sin^2(\bar{u}+z_i)}.
\eea
In appendix A we give the proof of this determinant representation of the partition functions.

%%%%%%%%%%%%%%%%%%%%%%%%%%%%%%%%%%%%%%%%%%%%%%%%%%%%%%%%%%%%%%%%%%%%%%%%%%%%%%%%%%%%%%%%%%%%%%%%%%%%%
\subsection{Scalar products $S^{1,1}$ and $S^{2,2}$}

In this subsection, we calculate the scalar products $S^{1,1}(\{u_\a\};\{v_i^{(1)}\})$ (\ref{ScalarBethe1})
and $S^{2,2}(\{u_\a\};\{v_i^{(2)}\})$ (\ref{ScalarBethe2}).
In this case, we assume that $\{v_i^{(k)}\}$ satisfy the associated Bethe ansatz equations.

Let us first consider the scalar product $S^{1,1}(\{u_\a\};\{v_i^{(1)}\})$ (\ref{ScalarBethe1}).
We insert in the scalar product a sum over the complete set of states
 $|j_1,\cdots,j_m\rangle$ between each operator,
 where $|j_1,\cdots,j_i\rangle$ is the state with $i$ spins down at the sites $j_1,\cdots,j_i$ and $2M-i$ spins up at the other sites.
  We are thus led to considering the intermediate functions
\bea
G^{(i)}(u_1,\cdots,u_i|j_{i+1},\cdots,j_M;\{v_i^{(1)}\})&=&
\langle j_{i+1},\cdots,j_M|\tilde{C}(u_i)\cdots\tilde{C}(u_1)\no\\
&&\times\tilde{B}(v_1^{(1)})\cdots\tilde{B}(v_M^{(1)})
|\Uparrow\rangle,
\label{intermediate}\eea
which satisfy the following recursive relations:
\bea
&&G^{(i)}(u_1,\cdots,u_i|j_{i+1},\cdots,j_M;\{v_i^{(1)}\})\no\\
&=&\sum_{j\neq j_{i+1},\cdots,j_M}
\langle j_{i+1},\cdots,j_M|\tilde{C}(u_i)|j,j_{i+1},\cdots,j_M\rangle\no\\
&&~~~\times G^{(i-1)}(u_1,\cdots,u_{i-1}|j,j_{i+1},\cdots,j_M;\{v_i^{(1)}\})
,~~~~i=1,\cdots,M.
\eea
The last one of these functions is the scalar product $S^{1,1}(\{u_\a\};\{v_i^{(1)}\})$, namely,
\bea
G^{(M)}(u_1,\cdots,u_M;\{v_i^{(1)}\})=S^{1,1}(\{u_\a\};\{v_i^{(1)}\}),
\eea
whereas the first one,
\bea
G^{(0)}(j_1,\cdots,j_M;\{v^{(1)}_i\})=\langle j_{1},\cdots,j_M|
\tilde{B}(v_1^{(1)})\cdots\tilde{B}(v_M^{(1)})
|\Uparrow\rangle
\eea
is closely related to the partition function $Z^{(2)}_{N}$ (\ref{partition function}) of the Gaudin model.

We then perform the summation in (\ref{intermediate}) and compute successively the functions $G^{(i)}$.
Details are given in Appendix B.
Finally, the scalar product $S^{1,1}(\{u_\a\};\{v_i^{(1)}\})$ has the following determinant representation
\bea
&&S^{1,1}_{M}(\{u_\a\},\{v_i^{(1)}\})=G^{(M)}(u_1,\cdots,u_M;\{v_i^{(1)}\})\no\\
&=&\frac{\det\widetilde{\mathscr{N}}^{1,1}(\{u_\a\};\{v_i^{(1)}\})}
{\prod_{k<j}\sin(u_k-u_j)\sin(u_k+u_j)\prod_{\a>\b}\sin(v^{(1)}_{\a}-v^{(1)}_{\b})
\sin(v^{(1)}_{\a}+v^{(1)}_{\b})},\label{scalar-product-1}
\eea
where the matrix $\widetilde{\mathscr{N}}^{1,1}(\{u_\a\};\{v_i^{(1)}\})$ is given by
\bea
\widetilde{\mathscr{N}}^{1,1}(\{u_\a\};\{v_i^{(1)}\})_{\a,j}=
\frac{\sin(2v^{(1)}_j)\sin(2u_{\a})F^{(1)}_j(u_\a;\{z_i\},\{v^{(1)}_i\})}
{\sin(\l_2+\x-v^{(1)}_j)\sin(\l_1+\x-v^{(1)}_j)}.
\eea
Here we have introduced the functions
\bea
&&F^{(1)}_j(u_\a;\{z_i\},\{v_i^{(1)}\})\no\\
\no\\&=&
\frac{\sin(\l_1+\x-u_{\a})\sin(\l_2+\x-u_{\a})\prod^M_{k=1}\sin(v^{(1)}_k-u_{\a})
\prod^M_{k=1}\sin(v^{(1)}_k+u_{\a})}
{\sin^2(v^{(1)}_j-u_{\a})\sin^2(v^{(1)}_j+u_{\a})}\no\\
&&\lt[\frac{1-\Delta}{\sin(\l_1+\x+u_{\a})\sin(\l_1+\x-u_{\a})}
+\frac{1+\Delta}{\sin(\l_2+\x+u_{\a})\sin(\l_2+\x-u_{\a})}\rt.\no\\
&&\lt.+\sum_{k=1}^{2M}\frac{1}{\sin(u_{\a}-z_k)\sin(u_{\a}+z_k)}
-\sum_{k\neq\a}^M\frac{2}{\sin(u_{\a}-v^{(1)}_k)\sin(u_{\a}+v^{(1)}_k)}\rt].
\eea

Similarly, we obtain the scalar product $S^{2,2}(\{u_\a\};\{v_i^{(2)}\})$ in terms of determinants
with the parameters $\{v^{(2)}_k\}$ satisfying the second set of Bethe ansatz equations (\ref{BA2}),
\bea
&&S^{2,2}(\{u_\a\};\{v_i^{(2)}\})\no\\
&=&\frac{\det\widetilde{\mathscr{N}}^{2,2}(\{u_\a\};\{v_i^{(2)}\})}
{\prod_{k<j}\sin(u_k-u_j)\sin(u_k+u_j)\prod_{\a>\b}\sin(v^{(2)}_{\a}-v^{(2)}_{\b})
\sin(v^{(2)}_{\a}+v^{(2)}_{\b})},
\eea
where the $M\times M$ matrix $\widetilde{\mathscr{N}}^{2,2}(\{u_\a\};\{v_i^{(2)}\})$ is given by
\bea
\widetilde{\mathscr{N}}^{2,2}(\{u_\a\};\{v_i^{(2)}\})_{\a,j}=
\frac{\sin(2v^{(2)}_j)\sin(2u_{\a})F^{(2)}_j(u_\a;\{z_i\},\{v^{(2)}_i\})}
{\sin(\l_2+\x+v^{(2)}_j)\sin(\l_1+\x+v^{(2)}_j)}.
\eea
Here
\bea
&&F^{(2)}_j(u_\a;\{z_i\},\{v_i^{(2)}\})\no\\
\no\\&=&
\frac{\sin(\l_1+\x+u_{\a})\sin(\l_2+\x+u_{\a})\prod^M_{k=1}\sin(v^{(2)}_k-u_{\a})
\prod^M_{k=1}\sin(v^{(2)}_k+u_{\a})}
{\sin^2(v^{(2)}_j-u_{\a})\sin^2(v^{(2)}_j+u_{\a})}\no\\
&&\lt[\frac{1+\Delta}{\sin(\l_1+\x+u_{\a})\sin(\l_1+\x-u_{\a})}
+\frac{1-\Delta}{\sin(\l_2+\x+u_{\a})\sin(\l_2+\x-u_{\a})}\rt.\no\\
&&\lt.+\sum_{k=1}^{2M}\frac{1}{\sin(u_{\a}-z_k)\sin(u_{\a}+z_k)}
-\sum_{k\neq\a}^M\frac{2}{\sin(u_{\a}-v^{(2)}_k)\sin(u_{\a}+v^{(2)}_k)}\rt]. \label{scalar-product-2}
\eea

\section{ Conclusions}
\label{C} \setcounter{equation}{0}

We have studied the XXZ Gaudin model with generic non-diagonal boundary terms. In addition to the inhomogeneous parameters
$\{z_j\}$, the associated Gaudin operators $\{H_j\}$, (\ref{Ham}), depend on four free parameters $\{\l_1,\,\l_2,\,\xi,\,\Delta\}$.
Thus our Gaudin operators are four-parameter ($\{\l_1,\,\l_2,\,\xi,\,\Delta\}$) generalizations of those in \cite{Ami01}, three-parameter
($\{\l_1,\,\l_2,\,\xi\}$)  generalizations of those in \cite{Hik95,Lor02},
and one-parameter ($\{\Delta\}$) generalizations of those in \cite{Yang04a}.
The common eigenstates (Bethe states) of the operators are constructed  by algebraic Bethe ansatz method.
We have obtained the determinant representations (\ref{partition function}),  (\ref{scalar-product-1}) and (\ref{scalar-product-2}) of
the scalar products for the boundary XXZ Gaudin model.

\section*{Acknowledgements}
The financial supports from  the National Natural Science
Foundation of China (Grant Nos. 10974247, 11075126, 11175248 and 11031005), the State Education Ministry of China
(Grant No. 20116101110017 and SRF for ROCS), the NWU Graduate Cross-discipline Fund (10YJC15) and
the Australian Research Council (Discovery Project DP11013434) are gratefully acknowledged.

\section*{Appendix A: Derivation of (\ref{partition function})}
\setcounter{equation}{0}
\renewcommand{\theequation}{A.\arabic{equation}}
Take the partition function $Z^{(2)}_N(\{v_{\a}\};\{z_i\})$ as an example, which satisfies the recursive relation,
\bea
Z^{(2)}_N(\{v_{\a}\};\{z_i\})&=&
\langle j_{1},\cdots,j_{N}|\tilde{B}(v_N)|j_{1},\cdots,j_{N-1}\rangle\no\\
&&~~~\times\langle j_{1},\cdots,j_{N-1}|
\tilde{B}(v_{N-1})\cdots\tilde{B}(v_1)
|\Uparrow\rangle,
\eea
namely
\bea
 Z^{(2)}_N(\{v_{\a}\};\{z_i\})&=&\sum_{i=1}^N
\frac{\sin(\l_1+\x+z_{i})\sin(\l_2+\x-z_{i})\sin(2v_N)}
{\sin(\l_1+\x-v_N)\sin(\l_2+\x-v_N)\sin(v_N-z_{i})
\sin(v_N+z_{i})}\no\\
\no\\
 &&\times Z^{(2)}_{N-1}(\{v_{\a}\}_{\a\neq N};\{z_j\}_{j\neq i}).\label{Recursive-relation}
\eea
The partition function $Z^{(2)}_N(\{v_{\a}\};\{z_i\})$, for any positive integer $N$,
can be uniquely determined by the initial condition:
$Z^{(2)}_0(\{v_{\a}\};\{z_i\})=1$ and the recursive relation (\ref{Recursive-relation}).

Let us introduce a series of functions $\{K_I(\{v_{\a}\};\{z_i\})|I=1,\cdots,N\}$,
\bea
&&K_N(\{v_{\a}\};\{z_i\})\no\\
&=&\frac{\prod^{N}_{\a=1}\prod^{N}_{i=1}\sin(v_{\a}+z_i)\sin(v_{\a}-z_i)}
{\prod_{\a>\b}\sin(v_{\a}-v_{\b})\sin(v_{\a}+v_{\b})\prod_{k<j}\sin(z_k-z_j)\sin(z_k+z_j)}\no\\
&&\times\det\left|\frac{\sin(\l_1+\x+z_j)\sin(\l_2+\x-z_j)\cdot2v_{\a}}
{\sin(\l_1+\x-v_{\a})\sin(\l_2+\x-v_{\a})\sin^2(v_{\a}-z_j)\sin^2(v_{\a}+z_j)}\right|,
\label{inductive function}\eea
and prove the relation,
\bea
Z^{(2)}_I(\{v_{\a}\};\{z_i\})=K_I(\{v_{\a}\};\{z_i\}),
\quad{\rm for\,any\,positive\,integer\,}I.
\label{partition function proof}
\eea
We use the induction method.
\begin{itemize}
\item We have, for the case of $N=1$,
\bea
Z^{(2)}_1(v_1;z_1)\hspace{-0.12truecm}=\hspace{-0.12truecm}K_1(v_1;z_1)
\hspace{-0.12truecm}=\hspace{-0.12truecm}
\frac{\sin(\l_1+\x+z_1)\sin(\l_2+\x-z_1)\sin(2v_1)}
{\sin(\l_1\hspace{-0.08truecm}+\hspace{-0.08truecm}\x\hspace{-0.08truecm}-\hspace{-0.08truecm}v_1)
\sin(\l_2\hspace{-0.08truecm}+\hspace{-0.08truecm}\x\hspace{-0.08truecm}-\hspace{-0.08truecm}v_1)
\sin(v_1\hspace{-0.08truecm}-\hspace{-0.08truecm}z_1)
\sin(v_1\hspace{-0.08truecm}+\hspace{-0.08truecm}z_1)}.
\eea
\item Suppose that (\ref{partition function proof}) holds for $I\leqslant N-1$.
We prove that (\ref{inductive function})  also holds for $I=N$ holds.
This can be done as follows. The determinant representation of $K_N(\{v_{\a}\};\{z_i\})$ implies that
it satisfies the following recursive relation
\bea
&&K_N(\{v_{\a}\};\{z_i\})\no\\
&=&\sum_{i=1}^N
\frac{\sin(\l_1+\x+z_{i})\sin(\l_2+\x-z_{i})\sin(2v_N)}
{\sin(\l_1+\x-v_N)\sin(\l_2+\x-v_N)\sin(v_N-z_{i})
\sin(v_N+z_{i})}\no\\
&&\times\prod^{N-1}_{l=1}\frac{\sin(v_l-z_i)\sin(v_l+z_i)}{\sin(v_N-v_l)\sin(v_N+v_l)}
\prod_{j\neq i}\frac{\sin(v_N-z_j)\sin(v_N+z_j)}{\sin(z_j-z_i)\sin(z_j+z_i)}\no\\
&&\times K_{N-1}(\{v_{\a}\}_{\a\neq N};\{z_j\}_{j\neq i}).
\label{inductive function recursive}\eea
The determinant representation (\ref{inductive function}), the recursive relation (\ref{inductive function recursive})
and the recursive relation (\ref{Recursive-relation}) mean that
$K_N(\{v_{\a}\};\{z_i\})$  and $Z^{(2)}_N(\{v_{\a}\};\{z_i\})$, as functions of $v_N$, have the same set of simple poles,
\bea
\pm z_i,\l_1+\x,\l_2+\x~~~~{\rm mod}(2\pi),~~i=1,\cdots,N,
\eea
at which both functions have the same residues. Moreover we can show that
\bea
Z^{(2)}_N(\{v_{\a}\};\{z_i\})|_{v_N\rightarrow\infty}=0=K_N(\{v_{\a}\};\{z_i\})
|_{v_N\rightarrow\infty}.\eea
We thus conclude that (\ref{partition function proof}) also holds for $I=N$. This completes the induction.
\end{itemize}
%Hence we have the polynomial representation,
%\bea
% Z^{(2)}_N(\{v_{\a}\};\{z_i\})&=&\sum_{\sigma\in{S_N}}\prod_{n=1}^N
%\frac{\sin(\l_1+\x+z_{i_{\s(n)}})\sin(\l_2+\x-z_{i_{\s(n)}})\sin(2v_n)}
%{\sin(\l_1+\x-v_n)\sin(\l_2+\x-v_n)\sin(v_n-z_{i_{\s(n)}})
%\sin(v_n+z_{i_{\s(n)}})}.\no\\
%\eea

%This fact allows us to prove
Finally we get the determinant representation of the partition function $Z^{(2)}_N(\{v_{\a}\};\{z_i\})$,
\bea
Z^{(2)}_N(\{v_\a\};\{z_i\})=
\frac{\prod^{N}_{\a=1}\prod^{N}_{i=1}\sin(v_{\a}+z_i)\sin(v_{\a}-z_i){\rm det}\mathscr{N}^{(2)}}
{\prod_{\a>\b}\sin(v_{\a}-v_{\b})\sin(v_{\a}+v_{\b})\prod_{k<j}\sin(z_k-z_j)\sin(z_k+z_j)},
\eea
where the $N\times N$ matrix $\mathscr{N}^{(2)}(\{v_{\a}\};\{z_j\})$ is given by
\bea
\mathscr{N}^{(2)}(\{v_{\a}\};\{z_j\})_{\a,j}=
\frac{\sin(\l_1+\x+z_j)\sin(\l_2+\x-z_j)\sin(2v_{\a})}
{\sin(\l_1\hspace{-0.08truecm}+\hspace{-0.08truecm}\x\hspace{-0.08truecm}-\hspace{-0.08truecm}v_{\a})
\sin(\l_2\hspace{-0.08truecm}+\hspace{-0.08truecm}\x\hspace{-0.08truecm}-\hspace{-0.08truecm}v_{\a})
\sin^2(v_{\a}\hspace{-0.08truecm}-\hspace{-0.08truecm}z_j)\sin^2(v_{\a}\hspace{-0.08truecm}+\hspace{-0.08truecm}z_j)}.
\eea

\section*{Appendix B: Proof of (\ref{scalar-product-1})}
\setcounter{equation}{0}
\renewcommand{\theequation}{B.\arabic{equation}}
We prove (\ref{scalar-product-1}) by calculating the functions $G^{(i)}$ (\ref{intermediate}) recursively.

%We calculate the intermediate functions $G^{(i)}$ (\ref{intermediate}) recursively in this Appendix.
%Briefly, the rational function proof we write here is only for the XXX Gaudin case, but the situation is quite similar for that of the XXZ Gaudin case.

We illustrate our derivations for $G^{(1)}$. We firstly express $G^{(1)}$ in terms of $G^{(0)}$,
\bea
&&\displaystyle G^{(1)}(u_1|j_2,\cdots,j_M;\{v^{(1)}_i\})\no\\\no\\
&=&\displaystyle\sum_{j\neq j_2,\cdots,j_M}
\langle j_2,\cdots,j_M|\tilde{C}(u_1)|j,j_2,\cdots,j_M\rangle
\times G^{(0)}(j,j_2,\cdots,j_M;\{v^{(1)}_i\})\no\\
 &=&\displaystyle\sum_{j\neq j_2,\cdots,j_M}
 \frac{\sin(\l_1+\x-z_i)\sin(\l_2+\x+z_i)\sin(2u_1)}{\sin(\l_1+\x+u_1)\sin(\l_2+\x+u_1)\sin(u_1-z_i)\sin(u_1+z_i)}\no\\
 &&~~~~\displaystyle\times
 \frac{\prod^{M}_{\a=1}\prod^{M}_{i=1}\sin(v^{(1)}_{\a}+z_i)\sin(v^{(1)}_{\a}-z_i){\rm det}\mathscr{N}(\{v^{(1)}_\a\};\{z_i\})}
{\prod_{\a>\b}\sin(v^{(1)}_{\a}-v^{(1)}_{\b})\sin(v^{(1)}_{\a}+v^{(1)}_{\b})\prod_{k<j}\sin(z_k-z_j)\sin(z_k+z_j)},
\eea
where $\mathscr{N}(\{v^{(1)}_\a\};\{z_i\})$ is a $M\times M$ matrix with matrix elements $\mathscr{N}(\{v^{(1)}_\a\};\{z_i\})_{\a,j}$ given by

\bea
\mathscr{N}(\{v^{(1)}_\a\};\{z_i\})_{\a,j}=
\frac{\sin(\l_1+\x+z_i)\sin(\l_2+\x-z_i)\sin(2v^{(1)}_{\a})}
{\sin(\l_1\hspace{-0.08truecm}+\hspace{-0.08truecm}\x\hspace{-0.08truecm}-\hspace{-0.08truecm}v^{(1)}_{\a})
\sin(\l_2\hspace{-0.08truecm}+\hspace{-0.08truecm}\x\hspace{-0.08truecm}-\hspace{-0.08truecm}v^{(1)}_{\a})
\sin^2(v^{(1)}_{\a}\hspace{-0.08truecm}-\hspace{-0.08truecm}z_i)
\sin^2(v^{(1)}_{\a}\hspace{-0.08truecm}+\hspace{-0.08truecm}z_i)}.
\eea
Further, performing the summation and absorbing the results into the element of the first column of the matrix $\mathscr{N}^{(1)}$ below,
we have
\bea
\displaystyle &&G^{(1)}(u_1|j_2,\cdots,j_M;\{v^{(1)}_i\})\no\\
&=&\displaystyle\frac{\prod^{M}_{\a=1}\prod^{M}_{k=2}\sin(v^{(1)}_{\a}+z_{i_k})\sin(v^{(1)}_{\a}-z_{i_k})
{\rm det}\mathscr{N}^{(1)}(\{v^{(1)}_{\a}\};u_1,i_2,\cdots,i_m)}
{\prod_{M\geq\a>\b\geq1}\sin(v^{(1)}_{\a}\hspace{-0.08truecm}-\hspace{-0.08truecm}v^{(1)}_{\b})
\sin(v^{(1)}_{\a}\hspace{-0.08truecm}+\hspace{-0.08truecm}v^{(1)}_{\b})
\prod_{2\leq k<l\leq \breve{} M}\sin(z_{i_k}\hspace{-0.08truecm}-\hspace{-0.08truecm}z_{i_l})
\sin(z_{i_k}\hspace{-0.08truecm}+\hspace{-0.08truecm}z_{i_l})},
\eea
where
\bea
\mathscr{N}^{(1)}_{ab}&=&\frac{1}{\sin(u_1-z_{i_b})\sin(u_1+z_{i_b})}\mathscr{N}_{ab}
~~~~~~~~~{\rm for}~~b\geq2,\\
\mathscr{N}^{(1)}_{a1}&=&
\displaystyle\prod^{M}_{k=2}\sin(u_1-z_{i_k})\sin(u_1+z_{i_k})\no\\
&&\times~~~~\hspace{-5mm}\sum^{M}_{i_1\neq i_2,\cdots,i_M}
\frac{\prod^{M}_{\a=1}\sin(v^{(1)}_\a+z_{i_1})\sin(v^{(1)}_\a-z_{i_1})}{\prod^{M}_{k>1}\sin(z_{i_1}-z_{i_k})\sin(z_{i_k}+z_{i_1})}\no\\
&&\times\displaystyle\frac{\sin(\l_1+\x+z_{i_1})\sin(\l_2+\x-z_{i_1})\sin(2v^{(1)}_a)}
{\sin(\l_1\hspace{-0.08truecm}+\hspace{-0.08truecm}\x\hspace{-0.08truecm}-\hspace{-0.08truecm}v^{(1)}_a)\sin(\l_2\hspace{-0.08truecm}+\hspace{-0.08truecm}\x\hspace{-0.08truecm}-\hspace{-0.08truecm}v^{(1)}_a)\sin^2(v^{(1)}_a\hspace{-0.08truecm}-\hspace{-0.08truecm}z_{i_1})\sin^2(v^{(1)}_a\hspace{-0.08truecm}+\hspace{-0.08truecm}z_{i_1})}\no\\
&&\times \frac{\sin(\l_1+\x-z_{i_1})\sin(\l_2+\x+z_{i_1})\sin(2u_1)}{\sin(\l_1\hspace{-0.08truecm}+\hspace{-0.08truecm}\x\hspace{-0.08truecm}+\hspace{-0.08truecm}u_1)\sin(\l_2\hspace{-0.08truecm}+\hspace{-0.08truecm}\x\hspace{-0.08truecm}+\hspace{-0.08truecm}u_1)\sin(u_1\hspace{-0.08truecm}-\hspace{-0.08truecm}z_{i_1})\sin(u_1\hspace{-0.08truecm}+\hspace{-0.08truecm}z_{i_1})}
,~~~~{\rm for}~~b=1.
\label{initial N1}\eea

Some remarks are in order. The summation in (\ref{initial N1}) is not over the full set of values, namely, $i_1\neq i_2\cdots i_M,$ (c.f. \cite{Kit99}).
Also there are second order poles (see (\ref{N1}) below), which make the calculation tedious.

Taking advantage of the analytical properties of the trigonometric function and keeping in mind that $\{v^{(1)}_i\}$
are solutions of the first set of Bethe ansatz equations, we can calculate  the sum in (\ref{initial N1}) to get
\bea
&&\hspace{-5mm}\sum^{M}_{i_1\neq i_2,\cdots,i_M}
\frac{\prod^{M}_{\a=1}\sin(v^{(1)}_\a+z_{i_1})\sin(v^{(1)}_\a-z_{i_1})}{\prod^{M}_{k>1}\sin(z_{i_1}-z_{i_k})\sin(z_{i_k}+z_{i_1})}\no\\
&&\times \frac{\sin(\l_1+\x-z_{i_1})\sin(\l_2+\x+z_{i_1})\sin(2u_1)}{\sin(\l_1+\x+u_1)\sin(\l_2+\x+u_1)\sin(u_1-z_{i_1})\sin(u_1+z_{i_1})}\no\\
&&\times\displaystyle\frac{\sin(\l_1+\x+z_{i_1})\sin(\l_2+\x-z_{i_1})\sin(2v^{(1)}_a)}
{\sin(\l_1+\x-v^{(1)}_a)\sin(\l_2+\x-v^{(1)}_a)\sin^2(v^{(1)}_a-z_{i_1})\sin^2(v^{(1)}_a+z_{i_1})}\no\\
&=&\frac{\widetilde{\mathscr{N}}_{a1}(u_1,\{v^{(1)}_\a\};\{z_{i_1}\})}{\prod^{M}_{k=2}\sin(u_1-z_{i_k})\sin(u_1+z_{i_k})}
+\sum^n_{b=2}\a_b{\mathscr{N}}^{(1)}_{ab}(u_1,\{v^{(1)}_\a\};\{z_{i_1}\}),
\label{N1}\eea
with
\bea
&&\widetilde{\mathscr{N}}_{a1}(u_1,\{v^{(1)}_\a\};\{z_i\})\no\\
&=&\hspace{-0.08truecm}\frac{\hspace{-0.08truecm}\sin(\l_1\hspace{-0.08truecm}+\hspace{-0.08truecm}\x\hspace{-0.08truecm}-\hspace{-0.08truecm}u_1)\hspace{-0.08truecm}
\sin(\l_2\hspace{-0.08truecm}+\hspace{-0.08truecm}\x\hspace{-0.08truecm}-\hspace{-0.08truecm}u_1)\hspace{-0.08truecm}\prod^M_{\a=1}\hspace{-0.08truecm}\sin(v^{(1)}_\a
\hspace{-0.08truecm}-\hspace{-0.08truecm}u_1)\hspace{-0.08truecm}\prod^M_{\a=1}\hspace{-0.08truecm}\sin(v^{(1)}_\a\hspace{-0.08truecm}+\hspace{-0.08truecm}u_1)
\hspace{-0.08truecm}\sin(2v^{(1)}_a)\hspace{-0.08truecm}\sin(2u_1)}
{\sin(\l_1+\x-v^{(1)}_a)\sin(\l_2+\x-v^{(1)}_a)\sin^2(v^{(1)}_a-u_1)\sin^2(v^{(1)}_a+u_1)}\no\\
&&\lt[\sum_{k=1}^{2M}\frac{1}{\sin(u_1-z_k)\sin(u_1+z_k)}-\sum_{k\neq\a}^M\frac{2}{\sin(u_1-v^{(1)}_k)\sin(u_1+v^{(1)}_k)}\rt.\no\\
&&\lt.+\frac{1-\Delta}{\sin(\l_1+\x+u_1)\sin(\l_1+\x-u_1)}
+\frac{1+\Delta}{\sin(\l_2+\x+u_1)\sin(\l_2+\x-u_1)}\rt],
\eea
where $\a_b$ are coefficients which do not depend on $a$. Then only the first term of the r.h.s. of (\ref{N1}) has nonzero contribution
to the determinant of $\widetilde{\mathscr{N}}$.

To show (\ref{N1}), we first prove the following relation:
\bea
&&\hspace{-5mm}\sum^{M}_{i_1\neq i_2,\cdots,i_M}
\frac{\prod^{M}_{\a=1}\sin(v^{(1)}_\a+z_{i_1})\sin(v^{(1)}_\a-z_{i_1})}{\prod^{M}_{k>1}\sin(z_{i_1}-z_{i_k})\sin(z_{i_k}+z_{i_1})}\no\\
&&\times \frac{\sin(\l_1+\x-z_{i_1})\sin(\l_2+\x+z_{i_1})\sin(2u_1)}{\sin(\l_1+\x+u_1)\sin(\l_2+\x+u_1)\sin(u_1-z_{i_1})\sin(u_1+z_{i_1})}\no\\
&&\times\displaystyle\frac{\sin(\l_1+\x+z_{i_1})\sin(\l_2+\x-z_{i_1})\sin(2v^{(1)}_a)}
{\sin(\l_1+\x-v^{(1)}_{a})\sin(\l_2+\x-v^{(1)}_{a})\sin^2(v^{(1)}_{a}-z_{i_1})\sin^2(v^{(1)}_{a}+z_{i_1})}\no\\
&=&\frac{\widetilde{\mathscr{N}}_{a1}\sin(u_1,\{v^{(1)}_\a\};\{z_i\})}
{\prod^{M}_{k=2}\sin(u_1-z_{i_k})\sin(u_1+z_{i_k})}-\frac{1}{\sin(\l_1+\x+u_1)\sin(\l_2+\x+u_1)}\no\\
&&\times\lt[\sum^{M}_{b=2}\frac{1}{\sin^2(u_1-z_{i_b})}
\frac{R(z_{i_b})}{\prod^{M}_{k=2,k\neq b}\sin(z_{i_b}-z_{i_k})\prod^{M}_{k=2}\sin(z_{i_b}+z_{i_k})}\rt.\no\\
&&+\sum^{M}_{b=2}\frac{1}{\sin(u_1-z_{i_b})}\lt\{
\frac{{\mathscr{N}}^{(1)}_{a,{i_b}}(u_1,\{v^{(1)}_\a\};\{z_i\})}
{\sin(\l_1+\x+z_{i_k})\sin(\l_2+\x-z_{i_k})}\rt.\no\\
&&\hspace{-5mm}\frac{\partial}{\partial u_1}\hspace{-0.08truecm}
\lt.\lt(\hspace{-0.08truecm}\frac{\prod^M_{\a=1}\hspace{-0.08truecm}\sin(v^{(1)}_\a\hspace{-0.08truecm}
-\hspace{-0.08truecm}u_1)\sin(v^{(1)}_\a\hspace{-0.08truecm}+\hspace{-0.08truecm}u_1)\sin(2u_1)
\hspace{-0.08truecm}\prod^2_{j=1}\hspace{-0.08truecm}\sin(\l_j\hspace{-0.08truecm}+\hspace{-0.08truecm}
\x\hspace{-0.08truecm}+\hspace{-0.08truecm}u_1)\sin(\l_j\hspace{-0.08truecm}+\hspace{-0.08truecm}\x\hspace{-0.08truecm}
-\hspace{-0.08truecm}u_1)}
{\prod^{M}_{k=2,k\neq{b}}\sin(u_1-z_{i_k})\sin(u_1+z_{i_k})}
\hspace{-0.08truecm}\rt)\hspace{-0.08truecm}\Bigg|_{u_1=z_{i_b}}\hspace{-0.08truecm}\rt\}
\no\\
&&+\sum^{M}_{b=2}\frac{1}{\sin(u_1-z_{i_b})}\frac{R(z_{i_b})\sin(2z_{i_b})}{\prod^{M}_{k=2,k\neq b}\sin(z_{i_b}-z_{i_k})\prod^{M}_{k=2}\sin(z_{i_b}+z_{i_k})}\no\\
&&\times\lt(\sum_{k=1}^{2M}\frac{1}{\sin(z_{i_b}-z_k)\sin(z_{i_b}+z_k)}\rt.
-\sum_{k=1}^M\frac{2}{\sin(z_{i_b}-v^{(1)}_k)\sin(z_{i_b}+v^{(1)}_k)}
\no\\
&&~~~~~+\frac{1-\Delta}{\sin(\l_1+\x+z_{i_b})\sin(\l_1+\x-z_{i_b})}
+\lt.\lt.\frac{1+\Delta}{\sin(\l_2+\x+z_{i_b})\sin(\l_2+\x-z_{i_b})}\rt)
\rt]\no\\
&&+~{\rm trigonometric~polynomials~with~poles~of}~\sin(u_1+z_{i_k}),
\label{check residues}\eea
where $R(z_{i_b})$ is
\bea
R(z_{i_b})&=&\sin(\l_1+\xi-z_{i_b})\sin(\l_2+\xi+z_{i_b})\prod^M_{\a=1}\sin(v^{(1)}_\a-z_{i_b})\prod^M_{\a=1}\sin(v^{(1)}_\a+z_{i_b})\sin(2z_{i_b})\no\\
&&\times\frac{\sin(\l_1+\xi+z_{i_b})\sin(\l_2+\xi-z_{i_b})\sin(2v^{(1)}_a)}
{\sin(\l_1+\xi-v^{(1)}_a)\sin(\l_2+\xi-v^{(1)}_a)\sin^2(v^{(1)}_a-z_{i_b})\sin^2(v^{(1)}_a+z_{i_b})}\no\\
&=&\sin(\l_1+\xi-z_{i_b})\sin(\l_2+\xi+z_{i_b})\prod^M_{\a=1}\sin(v^{(1)}_\a-z_{i_b})\prod^M_{\a=1}\sin(v^{(1)}_\a+z_{i_b})\sin(2z_{i_b})\no\\
&&\times{\mathscr{N}}^{(1)}_{a,{i_b}}(u_1,\{v^{(1)}_\a\};\{z_i\}).
\eea
Equation(\ref{check residues}) can be proven by considering both sides as meromorphic functions of $u_1$. Then both sides
of (\ref{check residues}) have the same set of simple poles {\footnote{We remark that the seemly apparent poles on the RHS of
(\ref{check residues}), located at $\pm z_{i_2},\ldots,z_{i_M}$, are actually not poles.}}:
\bea
 u_1=\pm z_{i_1},\,\,{\rm mod} (2\pi),\quad\quad {\rm where}\quad\quad i_1\neq i_2,\cdots,i_M,\no
\eea at which both have the same residues. Moreover,  both sides tend   $0$ as $u_1\longrightarrow\infty$.
We thus conclude that relation (\ref{check residues}) holds. Note that that  except the first one on the RHS of (\ref{check residues}) all other terms
 can be expressed in  the form of  $\sum_{b}\a_b{\mathscr{N}}^{(1)}_{ab}(u_1,\{v^{(1)}_\a\};\{z_i\})$. We thus obtain (\ref{N1}).

Repeating the above steps, by means of the recursive relations and changing the form of the determinant column by column,
we obtain the final expression for $S(\{u_\a\};\{v^{(1)}_i\})$.

%%%%%%%%%%%%%%%%%%%%%%%%%%%%%%%%%%%%%%%%%%%%%%%%%%%%%%%%%%%%%%%
%                                                             %
%  References                                                 %
%                                                             %
%%%%%%%%%%%%%%%%%%%%%%%%%%%%%%%%%%%%%%%%%%%%%%%%%%%%%%%%%%%%%%%

\end{document}